\documentclass[twocolumn,aps,pra,showpacs,preprintnumbers,amsmath,amssymb]{revtex4}
\usepackage{mathrsfs}

\usepackage{dcolumn}% Align table columns on decimal point
\usepackage{amsmath}
\usepackage{graphicx}% Include figure files
\usepackage{bm}% bold math
\usepackage{float}
\usepackage[dvipdfm,unicode,colorlinks,bookmarksnumbered=true,bookmarksopen=true,urlcolor=blue,CJKbookmarks=true]{hyperref}%生成中文书签和超链接

\begin{document}

\title{Large detuning limit for the multipartite systems interacting with electromagnetic fields}

\author{Hong-Yi Li}
 \email{hongyili@nudt.edu.cn}
 \author{Chun-Wang Wu}
\author{Ping-Xing Chen}
\author{Cheng-Zu Li}
\affiliation{College of Science, National University of Defense Technology, Changsha 410073, People's Republic of China}

\begin{abstract}
  We study the dynamics of the multipartite systems nonresonantly interacting with electromagnetic fields and show that, since the coupling strength $g$ is collectively enhanced by the square root of the number of microparticles involved, the more rigorous large detuning condition for neglecting the rapidly oscillating terms for the effective Hamiltonian should be $\Delta\gg\sqrt{N}g$, instead of $\Delta\gg g$ usually used in the literature even in the case of multipartite systems, with $\Delta$ the detuning and and $N$ the number of microparticles. This result is significant since merely the satisfaction of the original condition will result in the invalidity of the effective Hamiltonian and the errors of the parameters associated with the detuning in the multipartite case.\\

\end{abstract}

\maketitle

\section{Introduction}
Microparticles or artificial microparticles interacting with electromagnetic fields are important coupling models to accomplish various quantum information processing tasks, such as quantum gate \cite{Rabl,DiCarlo,Gulde}, quantum memory \cite{Choi,Zhao} and quantum communication \cite{Duan,Han}. Many efforts have been devoted to the investigation of these physical processes theoretically \cite{James,Gamel,Brion}.

Recently, James and Jerke presented a useful compact formula for deriving an effective Hamiltonian describing the time-averaged dynamics of detuned quantum systems \cite{James}. Then Gamel and James developed a technique for finding the dynamical evolution  of the time-averaged density matrix, and the equation of evolution includes an effective Hamiltonian, as well as decoherence terms in Lindblad form \cite{Gamel}. Brion \textit{et al.} derived  an unambiguous effective two-level Hamiltonian of a lambda system excited by off-resonant laser beams and quantified the accuracy of the approximation achieved \cite{Brion}. In this paper, we focus on the large detuning condition for neglecting the rapidly oscillating terms for the effective Hamiltonians of the multipartite systems interacting with the electromagnetic fields.

In the large detuning regime, the energy exchanges between the field modes and the relevant microparticle transitions are suppressed and the rapidly oscillating terms in the original Hamiltonian could be neglected approximately to get a simpler effective Hamiltonian, which no longer contains the energy exchange terms between the field modes and the relevant microparticle transitions accordingly. The large detuning condition is usually formulated as $\Delta\gg g$ even in the case of multipartite systems, actually, which is sufficient only for few microparticles involved. When a large number of microparticles interact with the fields, the coupling strength $g$ could be collectively enhanced due to the many-particle interference effects \cite{Duan,Gorshkov}, therefore the large detuning condition in multipartite situation should scale with the number of microparticles involved accordingly. We show that the rigorous large detuning condition should be expressed as $\Delta\gg\sqrt{N}g$ in the multipartite case. In fact, if merely $\Delta\gg g$ is satisfied, there may exist notable real energy exchanges between the field modes and the relevant microparticle transitions in multipartite case, however the effective Hamiltonian contains no energy exchange terms between them, meaning its invalidity of describing the coupled system. In mathematical language, $\Delta\gg g$ is necessary but not sufficient for the validity of the effective Hamiltonians of the multipartite systems interacting with fields, while $\Delta\gg\sqrt{N}g$ is just necessary and sufficient for it.  Although by the full numerical simulations for this type of problem, one may observe that the detuning should be chosen much larger than the value determined by the condition $\Delta\gg g$, this is a time-consuming task \cite{Xiao}. Furthermore, for the theoretical works without full numerical simulations, the invalidity of the effective Hamiltonians resulting from the original large detuning condition may not be realized \cite{Dasgupta,Wu,Lee,Lv}, and the incorrect detuning value will lead to other wrong parameters associated with it, such as the gating time.

In what follows, we take the cavity QED system as example to derive the large detuning condition for the effective Hamiltonian of many atoms interacting with fields. First we calculate the exact solution of the time-evolution operator for the two-level atoms nonresonantly interacting with the cavity mode and numerically simulate the state evolution. Then we analyze the cases of three-level $\Lambda$ atoms off-resonantly interacting with the cavity mode and the classical field under the conditions of two-photon resonance and nonresonace.

\section{Two-Level Configuration}

First, let us consider an interacting system consisting of an ensemble of atoms with two-level configuration nonresonantly coupled to a single-mode cavity field. The atoms are assumed for simplicity to have the same coupling strength $g$ and detuning $\Delta$. In the interaction picture, the coupled system can be described by the Hamiltonian (assuming $\hbar =1$)
\begin{equation}\label{eq1}
H=\sum\limits_{j=1}^{N}{g\left( \sigma _{j}^{-}{{a}^{\dag}}{{e}^{-i\Delta t}}+\sigma _{j}^{\dag}a{{e}^{i\Delta t}} \right)},
\end{equation}
where the subscript $j$ represents the $j$th atom, $a$ $(a^{\dag})$ denotes the annihilation (creation) operator for the cavity mode, $N$ is the number of atoms involved and $\sigma _{j}^{-}=|g_{j}\rangle\langle e_{j}|$, $\sigma _{j}^{\dag}=|e_{j}\rangle\langle g_{j}|$ with $|e_{j}\rangle$, $|g_{j}\rangle$ ($j=1,2,\cdots,N$) being the excited and ground states of the $j$th atom. For suppressing the cavity decay, we require no real energy exchange between the cavity mode and the atomic ensemble, and the cavity mode in vacuum state initially. To this end the coupled system need to work in a large detuning regime usually expressed in the form of $\Delta\gg g$ in literatures even in the case of a large number of atoms involved \cite{Dasgupta,Lee}. While we will show this is not sufficient to prevent the real energy exchanging between the cavity and the multi-atomic system.

In the context of quantum information processing with microparticle ensembles \cite{Rabl,Chen,Tanji,Duan}, one usually considers the singly excited situation, i.e., only one atom is excited in the ensemble while the other atoms in the ground states and the cavity mode in the vacuum state, or the cavity mode is in the single-photon state while all the atoms in the ground states. In this case there exists an invariant subspace spanned by $\{|+\rangle=|E\rangle|0\rangle,|-\rangle=|G\rangle|1\rangle\}$ for the system evolution, where $|E\rangle=(1/\sqrt{N})\sum\nolimits_{j=1}^{N}|g_1 \cdots g_{j-1} e_j  g_{j+1}\cdots g_N\rangle$, $|G\rangle=|g_1 g_2 \cdots g_N\rangle$, and $|0\rangle$ ($|1\rangle$) denotes the cavity mode in the vacuum state (in the single-photon state). We may call $|E\rangle$ the collective atomic excited state and $|G\rangle$ the collective atomic ground state \cite{Riedmatten,Choi}. In terms of collective atomic operators, the Hamiltonian (\ref{eq1}) can be written as
\begin{equation}
    H=\sqrt{N}gS{{a}^{\dag}}{{e}^{-i\Delta t}}+\sqrt{N}g{{S}^{\dag}}a{{e}^{i\Delta t}},
\end{equation}
\begin{figure}
\includegraphics[scale=0.5]{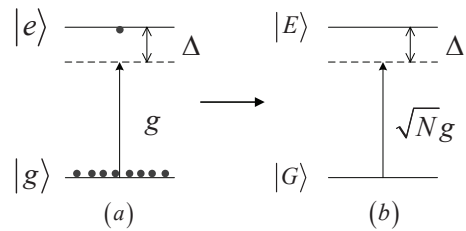}
\caption{\label{fig1}
The collective atomic level diagram for the two-level configuration. (a) Considering the singly excited situation. The single-atom-cavity coupling constant is $g$ and the detuning between single atomic resonance and cavity mode is $\Delta$. (b) The coupling strength
between the collective atomic states and the cavity mode is $\sqrt{N}$ times larger than the single-atom-cavity coupling
constant, however the detuning between the collective atomic resonance and the cavity mode is still $\Delta$.}
\end{figure}%
\begin{figure}
\includegraphics[scale=0.65]{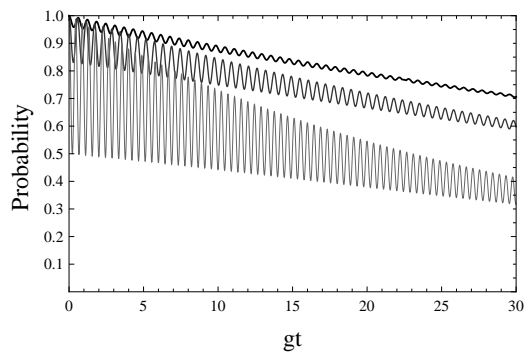}
\caption{\label{fig2}
Temporal evolution of the  state $|+\rangle$ under atom and cavity decays with the initial state in $|+\rangle$. The three solid curves from top to bottom correspond to the number of atoms in the cavity being 1, 5, and 25, respectively. Other parameters: $\Delta=10g$, $\kappa=g/10$, and $\gamma=g/100$, where $\kappa$ is the cavity decay rate and $\gamma$ is the atomic spontaneous emission rate.
}
\end{figure}%
where $S=|G\rangle\langle E|$ and ${S}^{\dag}=|E\rangle\langle G|$ are collective atomic lowering and raising operators. We can see that the coupling constant is enhanced by the square root of the number of atoms involved, however the detuning is still $\Delta$. From this Hamiltonian, as well as the collective atomic level diagram shown in Fig.~\ref{fig1}, if we expect there is no real energy exchange between the cavity mode and the atomic ensemble, then $\Delta\gg \sqrt{N}g$ is required for the large detuning regime, instead of the original condition $\Delta\gg g$. In the following, we give a rigorous derivation for this result through solving the exact solution of the temporal evolution of the coupled system.

After a rotating transformation with respect to $-\Delta\sum\nolimits_{j=1}^{N}\sigma_{j}^{z}/2$ with $\sigma_{j}^{z}=|e_{j}\rangle\langle e_{j}|-|g_{j}\rangle\langle g_{j}|$, the Hamiltonian (\ref{eq1}) goes into
\begin{equation}
   H_{R}=\sum\limits_{j=1}^{N}{\left[ \frac{\Delta }{2}\sigma _{j}^{z}+g\left( \sigma _{j}^{-}{{a}^{+}}+\sigma _{j}^{+}a \right) \right]}.
\end{equation}
In the subspace spanned by $\{|+\rangle=|E\rangle|0\rangle,|-\rangle=|G\rangle|1\rangle\}$, the Hamiltonian $H_R$ can be written as
\begin{equation}
  {{H}_{R}}={{H}_{1}}+{{H}_{2}},
\end{equation}
with
\begin{equation}
  {{H}_{1}}=-\frac{\left( N-1 \right)\Delta }{2}I ,~~~~
  {{H}_{2}}=\left[ \begin{matrix}
   \frac{\Delta }{2} & g\sqrt{N}  \\
   g\sqrt{N} & -\frac{\Delta }{2}
\end{matrix} \right],
\end{equation}
where $I$ is the $2\times2$ identity matrix. The time-evolution operator of the system can be written as
 \begin{equation}\label{}
    U(t)=e^{-itH_1}e^{-itH_2}.
 \end{equation}
Neglecting the global phase factor resulting from $H_1$, through direct calculation the exact solution of the time-evolution operator $U(t)$ is given by
\begin{equation}
U\left( t \right)=\left[
\begin{matrix}
   A_{\alpha }^{-} & {{B}_{\beta }}  \\
   {{B}_{\beta }} & A_{\alpha }^{+}  \\
\end{matrix}
\right],
\end{equation}
where
\begin{eqnarray}
  & A_{\alpha }^{-}=\cos (\frac{1}{2}\alpha \Delta t)-\frac{i}{\alpha }\sin (\frac{1}{2}\alpha \Delta t),\nonumber
   \\
 & A_{\alpha }^{+}=\cos (\frac{1}{2}\alpha \Delta t)+\frac{i}{\alpha }\sin (\frac{1}{2}\alpha \Delta t) , \nonumber
 \\
 & {{B}_{\beta }}=-i\beta \sin \left( \frac{1}{2}\alpha \Delta t \right),\nonumber
  \\
 & \alpha =\sqrt{1+{{\theta }^{2}}},~~\beta =\sqrt{\frac{1}{1+1/{{\theta }^{2}}}},~~\theta =\frac{2\sqrt{N}g}{\Delta }.\label{eq}
\end{eqnarray}
In the large detuning regime the effective Hamiltonian is given by \cite{Gamel}
\begin{equation}\label{eqh}
    H_{eff}=\lambda\sum_{j=1}^{N}\left( \left|e_{j}\right\rangle\left\langle e_{j}\right|a a^{\dag}-\left|g_{j}\right\rangle\left\langle g_{j}\right| a^{\dag} a \right)+\lambda\sum_{\substack{j,k=1\\j\neq k}}^{N}\sigma_{j}^{\dag}\sigma_{k}^{-},
\end{equation}
where $\lambda=g^{2}/\Delta$. From Eq.~(\ref{eqh}), we see that there is no real energy exchange between the cavity mode and the atoms, thus the off-diagonal elements ${B}_{\beta }$ in the temporal evolution operator should approximate to zero all the time, and according to Eqs.~(\ref{eq}) we obtain $\Delta\gg \sqrt{N}g$, in accordance with the physical picture shown in Fig.~\ref{fig1}. With numerical simulation employing master equation, figure~\ref{fig2} shows the time evolution of the state $|+\rangle$ under atom and cavity decays assuming the initial state in $|+\rangle$. We can see that, when the number of atoms in the cavity grows, the system is not only more likely to go out of the state $|+\rangle$ but also decays faster, which both indicate explicitly that the original large detuning condition become inefficient gradually with atom number increasing.

\section{Three-Level $\Lambda$ Configuration}
\begin{figure}
\includegraphics[scale=0.42]{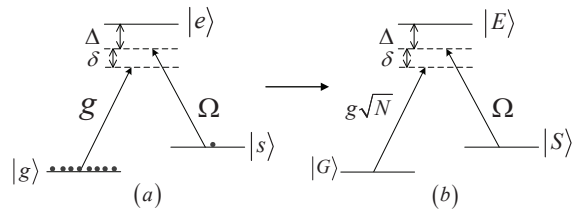}
\caption{\label{fig3}
The collective atomic level diagram for the $\Lambda$-level configuration. (a) Considering
the singly excited situation. The single-atom-cavity coupling constant is $g$ with detuning $\Delta+\delta$
and the coupling strength between the single atom and the classical field is $\Omega$ with detuning $\Delta$.
(b) The coupling strength between the cavity mode and atoms is enhanced by the square root of the number of
atoms involved, while the detuning is still $\Delta+\delta$.}
\end{figure}

Now, let us consider an ensemble of atoms with three-level $\Lambda$ configuration (shown in Fig.~\ref{fig3}) simultaneously interacting with a single-mode cavity and driven by a classical field. The two lower states $|g\rangle$ and $|s\rangle$ can be achieved, for example, with hyperfine or Zeeman sublevels of electronic ground states, and $|e\rangle$ is the excited state. The $\left| g \right\rangle \leftrightarrow \left| e \right\rangle $ transition is nonresonantly coupled to the cavity mode with the atom-cavity coupling strength $g$ and detuning $\Delta+\delta$, while the $\left| s \right\rangle \leftrightarrow \left| e \right\rangle $ transition is nonresonantly driven by the classical field with the Rabi frequency $\Omega $ and detuning $\Delta $. In the interaction picture, the coupled system can be described by the Hamiltonian (assuming $\hbar =1$)
\begin{equation}\label{eq2}
 H=\sum\limits_{j=1}^{N}{\left( g\left| {{e}_{j}} \right\rangle \left\langle  {{g}_{j}} \right|a{{e}^{i(\Delta+\delta)t}}+\Omega \left| {{e}_{j}} \right\rangle \left\langle  {{s}_{j}} \right|{{e}^{i\Delta t}}+H.c. \right)}.
\end{equation}
Consider the singly excited situation similar as in the case of two-level configuration, then the system evolution will be restricted in the Hilbert space spanned by $\{ \left| {{u}_{1}} \right\rangle =\left| E \right\rangle \left| 0 \right\rangle$, $\left| {{u}_{2}} \right\rangle =\left| G \right\rangle \left| 1 \right\rangle$, $\left| {{u}_{3}} \right\rangle =\left| S \right\rangle \left| 0 \right\rangle \}$, where $|S\rangle=(1/\sqrt{N})\sum\nolimits_{j=1}^{N}|g_1 \cdots g_{j-1} s_j  g_{j+1}\cdots g_N\rangle$. In terms of collective atomic operators, the Hamiltonian (\ref{eq2}) can be written as
\begin{equation}
 H= \sqrt{N}g \mathrm{S}_{G}^{\dag} a{{e}^{i(\Delta+\delta)t}}+\Omega \mathrm{S}_{S}^{\dag}{{e}^{i\Delta t}}+H.c. ,
\end{equation}
where $\mathrm{S}_{G}^{\dag}=\left| E \right\rangle \left\langle  G \right|$ and $\mathrm{S}_{S}^{\dag}=\left| E \right\rangle \left\langle  S \right|$. We can see that the effective coupling strength between the atoms and the cavity mode is $\sqrt{N}$ times larger than the single-atom-cavity coupling constant, while the detuning between the $\left| G \right\rangle \leftrightarrow \left| E \right\rangle $ transition and the cavity mode is still $\Delta+\delta$ (as shown in Fig.~\ref{fig3}). Then the excited state $\left| e \right\rangle$ of short lifetime could be adiabatically eliminated during the operations, if the excited state is not occupied initially and $\Delta \gg \sqrt{N}g, \Omega$ ($\Delta \gg g$ is not sufficient). In what follows, we prove this result in two situations: two-photon resonance and nonresonance, and give the condition for eliminating the cavity mode in the latter case.

\subsection{Two-photon resonance situation}
First, let us consider the simple case of two-photon resonance, i.e., $\delta=0$. After a rotating transformation \cite{Brion} with respect to $-\Delta \sum_{j=1}^{N}{\left| {{e}_{j}} \right\rangle \left\langle  {{e}_{j}} \right|}$, the Hamiltonian (\ref{eq2}) goes into
\begin{equation}
   H_R=\sum\limits_{j=1}^{N}{\left[ \Delta \left| {{e}_{j}} \right\rangle \left\langle  {{e}_{j}} \right|+\left( g\left| {{e}_{j}} \right\rangle \left\langle  {{g}_{j}} \right|a+\Omega \left| {{e}_{j}} \right\rangle \left\langle  {{s}_{j}} \right|+H.c. \right) \right]}.
\end{equation}
In the subspace spanned by $\{ \left| {{u}_{1}} \right\rangle =\left| E \right\rangle \left| 0 \right\rangle$, $\left| {{u}_{2}} \right\rangle =\left| G \right\rangle \left| 1 \right\rangle$, $\left| {{u}_{3}} \right\rangle =\left| S \right\rangle \left| 0 \right\rangle \}$, the Hamiltonian $H_R$ can be written as
\begin{equation}
   {{H}_{R}}=\left[
   \begin{matrix}
   \Delta  & \sqrt{N}g & \Omega   \\
   \sqrt{N}g & 0 & 0  \\
   \Omega  & 0 & 0  \\
\end{matrix}
\right].
\end{equation}
The exact solution of the temporal evolution operator up to a global phase factor $e^{-it\Delta/2}$ is given as follows:
\begin{equation}
    U\left( t \right)=\left[
     \begin{matrix}
   A_{\alpha }^{-} & {{B}_{\beta }} & {{B}_{\gamma }}  \\
   {{B}_{\beta }} & {{D}_{1/\eta }} & {{C}_{\eta }}  \\
   {{B}_{\gamma }} & {{C}_{\eta }} & {{D}_{\eta }}  \\
\end{matrix}
\right],
\end{equation}
where
\begin{eqnarray}
  & {{C}_{\eta }}=-\frac{1}{\eta +1/\eta }{{e}^{i\Delta t/2}}+\frac{1}{\eta +1/\eta }A_{\alpha }^{+}, \nonumber\\
 & {{D}_{\eta }}=\frac{1}{1+{{\eta }^{2}}}{{e}^{i\Delta t/2}}+\frac{1}{1+1/{{\eta }^{2}}}A_{\alpha }^{+}, \nonumber\\
 & {{D}_{1/\eta }}=\frac{1}{1+1/{{\eta }^{2}}}{{e}^{i\Delta t/2}}+\frac{1}{1+{{\eta }^{2}}}A_{\alpha }^{+},~~\eta =\sqrt{N}g/\Omega,  \nonumber\\
 & A_{\alpha }^{\pm }=\cos \left( \frac{1}{2}\Delta \alpha t \right)\pm \frac{i}{\alpha }\sin \left( \frac{1}{2}\Delta \alpha t \right), \nonumber\\
 & {{B}_{\beta }}=-\beta \sin \left( \frac{1}{2}\Delta \alpha t \right),~~{{B}_{\gamma }}=-\gamma \sin \left( \frac{1}{2}\Delta \alpha t \right), \nonumber\\
 & \alpha =\sqrt{4N{{g}^{2}}+4{{\Omega }^{2}}+{{\Delta }^{2}}}/\Delta , \nonumber\\
 & \beta =2\sqrt{N}g/\sqrt{4N{{g}^{2}}+4{{\Omega }^{2}}+{{\Delta }^{2}}}, \nonumber\\
 & \gamma =2\Omega /\sqrt{4N{{g}^{2}}+4{{\Omega }^{2}}+{{\Delta }^{2}}}. \label{eq3}
\end{eqnarray}
\begin{figure}
\includegraphics[scale=0.58]{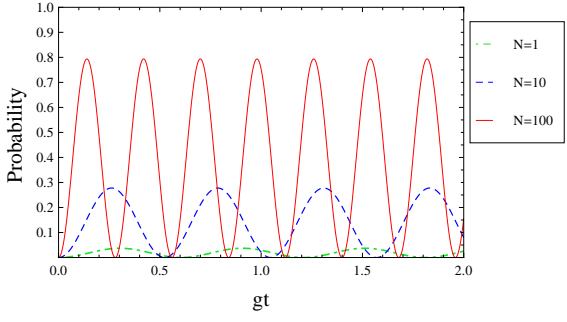}
\caption{\label{fig4}(Color online) Temporal evolution of the  state $|u_{1}\rangle=\left| E \right\rangle \left| 0 \right\rangle$ with the initial state in $|u_{2}\rangle=\left| G \right\rangle \left| 1 \right\rangle$ considering no dissipations. With the number of atoms in the cavity growing, there exists more real energy exchanges between the cavity mode and $\left| g \right\rangle \leftrightarrow \left| e \right\rangle $ transition in a certain detuning $\Delta$. Other parameters: $\Omega=g$ and $\Delta=10g$ only enough for few atoms involved to prevent the $\left| g \right\rangle \leftrightarrow \left| e \right\rangle $ transition.
}
\end{figure}%
In the large detuning regime, we can eliminate the excited state adiabatically, that means the transitions between the lower states and the excited state are forbidden, thus ${B}_{\beta }$ and ${B}_{\gamma }$ should approximate to zero all the time and according to Eqs.~(\ref{eq3}) $\Delta\gg\sqrt{N}g,\Omega$ should be satisfied. Fig.~\ref{fig4} shows that, as the number of atoms involved grows, the excited states not occupied initially  are populated with higher probability during the state evolution in a certain detuning $\Delta$, which reveals that the large detuning condition should be relevant to the number of atoms involved.

It should be pointed out that there may exist real energy exchanges between the cavity mode and atoms even in the large enough detuning due to the two-photon resonance, however it is between the cavity mode and the $\left| g \right\rangle \leftrightarrow \left| s \right\rangle $ transition, not between the cavity mode and the $\left| g \right\rangle \leftrightarrow \left| e \right\rangle $ transition. In the following, we analyze the two-photon nonresonance situation, where if the detuning $\delta$ is large enough, the real energy exchanges between the cavity mode and the $\left| g \right\rangle \leftrightarrow \left| s \right\rangle $ transition could also be blocked.

\subsection{Two-photon nonresonance situation}
In this case ($\delta \neq 0$), we perform a rotating transformation with respect to $\sum\nolimits_{j=1}^{N}{\left[ \left( \delta +\Delta  \right)\left| {{g}_{j}} \right\rangle \left\langle  {{g}_{j}} \right|+\Delta \left| {{s}_{j}} \right\rangle \left\langle  {{s}_{j}} \right| \right]}$, the Hamiltonian (\ref{eq2}) goes into
\begin{align}
    {{H}_{R}}=&\sum\limits_{j=1}^{N}{\left[ -\left( \delta +\Delta  \right)\left| {{g}_{j}} \right\rangle \left\langle  {{g}_{j}} \right|-\Delta \left| {{s}_{j}} \right\rangle \left\langle  {{s}_{j}} \right| \right.}\nonumber \\
    &\left.+\left( g\left| {{e}_{j}} \right\rangle \left\langle  {{g}_{j}} \right|a+\Omega \left| {{e}_{j}} \right\rangle \left\langle  {{s}_{j}} \right|+H.c. \right) \right].
\end{align}
In the subspace spanned by $\{ \left| {{u}_{1}} \right\rangle =\left| E \right\rangle \left| 0 \right\rangle$, $\left| {{u}_{2}} \right\rangle =\left| G \right\rangle \left| 1 \right\rangle$, $\left| {{u}_{3}} \right\rangle =\left| S \right\rangle \left| 0 \right\rangle \}$, the Hamiltonian $H_R$ can be written as
\begin{equation}
  {{H}_{R}}={{H}_{1}}+{{H}_{2}},
\end{equation}
with
\begin{equation}
   {{H}_{1}}=-N\left( \delta +\Delta  \right)I,~~~{{H}_{2}}=\left[ \begin{matrix}
   \delta +\Delta  & \sqrt{N}g & \Omega   \\
   \sqrt{N}g & 0 & 0  \\
   \Omega  & 0 & \delta   \\
\end{matrix} \right],
\end{equation}
where $I$ is the $3\times3$ identity matrix, and ${H}_{1}$ only results in a global phase factor, thus we could neglect it in the derivation of the state evolution. Assuming the initial state $\left| \varphi (t=0) \right\rangle=\alpha_{0}\left| {{u}_{1}} \right\rangle+\beta_{0}\left| {{u}_{2}} \right\rangle+\gamma_{0}\left| {{u}_{3}} \right\rangle$, the derivation of the exact solution of the state evolution can be performed by finding the eigenenergies of the system and using the initial condition to determine the coefficients of the Fourier decompositions of the different amplitudes. Since the exact solution of the state evolution is too complicated to write down and analyze, we introduce a set of reduced variables ($\lambda,\lambda_{c},\lambda_{l}$) such that $\lambda  \epsilon ={\delta }/{\Delta }, \lambda _c \epsilon ={\sqrt{N} g}/{\Delta }, \lambda _l \epsilon ={\Omega }/{\Delta }$, with $0<\epsilon<1$, $(\lambda,\lambda_{c},\lambda_{l})=O(1)$, to simplify the expression of the state evolution but preserving the most important components. We do not reproduce the calculations but only give the final result of the state evolution $\left| \varphi (t) \right\rangle=\alpha(t)\left| {{u}_{1}} \right\rangle+\beta(t)\left| {{u}_{2}} \right\rangle+\gamma(t)\left| {{u}_{3}} \right\rangle$, with
\begin{align}
  \alpha(t)=& \sum_{j=1}^{3}\alpha_{j}e^{-it\Delta p_{j}}, \nonumber\\
  \beta(t)=& \sum_{j=1}^{3}\beta_{j}e^{-it\Delta p_{j}}, \nonumber\\
  \gamma(t)=& \sum_{j=1}^{3}\gamma_{j}e^{-it\Delta p_{j}},
\end{align}
where $\{p_{1},p_{2},p_{3}\}$ are the solutions of the equation
\begin{equation}
    p^3- (2 \lambda  \epsilon +1)p^2- \left( \lambda _c^2\epsilon ^2+ \lambda _l^2\epsilon ^2-\lambda ^2\epsilon  ^2- \lambda \epsilon \right)p+ \lambda  \lambda _c^2\epsilon ^3=0,
\end{equation}
and the coefficients $\{\alpha_{j},\beta_{j},\gamma_{j}\}_{j=1,2,3}$ are determined by the initial condition. The expansions in $\epsilon$ of these different parameters are given by
\begin{align}
  p_{1} =&1+ \lambda \epsilon +O\left(\epsilon ^2\right),~p_{2}=O\left(\epsilon ^2\right),~p_{3}=\lambda \epsilon +O\left(\epsilon ^2\right), \nonumber \\
  \alpha_{1} =& \alpha _0+\left(\beta _0 \lambda _c+\gamma _0 \lambda _l\right) \epsilon +O\left(\epsilon ^2\right),\nonumber \\
  \alpha_{2} =&-\beta _0 \lambda _c \epsilon +O\left(\epsilon ^2\right),\nonumber \\
  \alpha_{3} =&-\gamma _0 \lambda _l \epsilon +O\left(\epsilon ^2\right),\nonumber \\
  \beta_{1}  =&\alpha _0 \lambda _c \epsilon +O\left(\epsilon ^2\right),\nonumber \\
  \beta_{2}  =&\beta _0+\frac{\lambda _c \left(\gamma _0 \lambda _l-\lambda  \alpha _0\right) \epsilon }{\lambda }+O\left(\epsilon ^2\right),\nonumber \\
  \beta_{3}  =&-\frac{\gamma _0 \lambda _c \lambda _l \epsilon }{\lambda }+O\left(\epsilon ^2\right),\nonumber \\
  \gamma_{1} =&\alpha _0 \lambda _l \epsilon +O\left(\epsilon ^2\right),\nonumber \\
  \gamma_{2} =&\frac{\beta _0 \lambda _c \lambda _l \epsilon }{\lambda }+O\left(\epsilon ^2\right),\nonumber \\
  \gamma_{3} =&\gamma _0-\frac{\left(\lambda  \alpha _0+\beta _0 \lambda _c\right) \lambda _l \epsilon }{\lambda }+O\left(\epsilon ^2\right).
\end{align}
Assuming the initial state is $\left| {{u}_{2}} \right\rangle =\left| G \right\rangle \left| 1 \right\rangle$, i.e., $(\alpha_{0}=0,\beta_{0}=1,\gamma_{0}=0)$, then the solution of the state evolution $\left| \varphi (t) \right\rangle=\alpha(t)\left| {{u}_{1}} \right\rangle+\beta(t)\left| {{u}_{2}} \right\rangle+\gamma(t)\left| {{u}_{3}} \right\rangle$ can be obtained from the coefficients:
\begin{align}
  \alpha(t) =&O\left(\epsilon ^2\right)e^{-it\Delta p_{3}}+\left[-\frac{\sqrt{N} g}{\Delta } +O\left(\epsilon ^2\right)\right]e^{-it\Delta p_{2}} \nonumber \\
  &+\left[\frac{\sqrt{N} g}{\Delta } +O\left(\epsilon ^2\right)\right]e^{-it\Delta p_{1}},\nonumber \\
  \beta(t) =&O\left(\epsilon ^2\right)e^{-it\Delta p_{1}}+\left[1+O\left(\epsilon ^2\right)\right]e^{-it\Delta p_{2}} \nonumber \\
  &+O\left(\epsilon ^2\right)e^{-it\Delta p_{3}},\nonumber \\
  \gamma(t) =&O\left(\epsilon ^2\right)e^{-it\Delta p_{1}}+\left[\frac{\frac{\sqrt{N}g\Omega}{\Delta}}{\delta}+O\left(\epsilon ^2\right)\right]e^{-it\Delta p_{2}} \nonumber \\
  &+\left[-\frac{\frac{\sqrt{N}g\Omega}{\Delta}}{\delta}+O\left(\epsilon ^2\right)\right]e^{-it\Delta p_{3}}.
\end{align}
\begin{figure}
\includegraphics[scale=0.4]{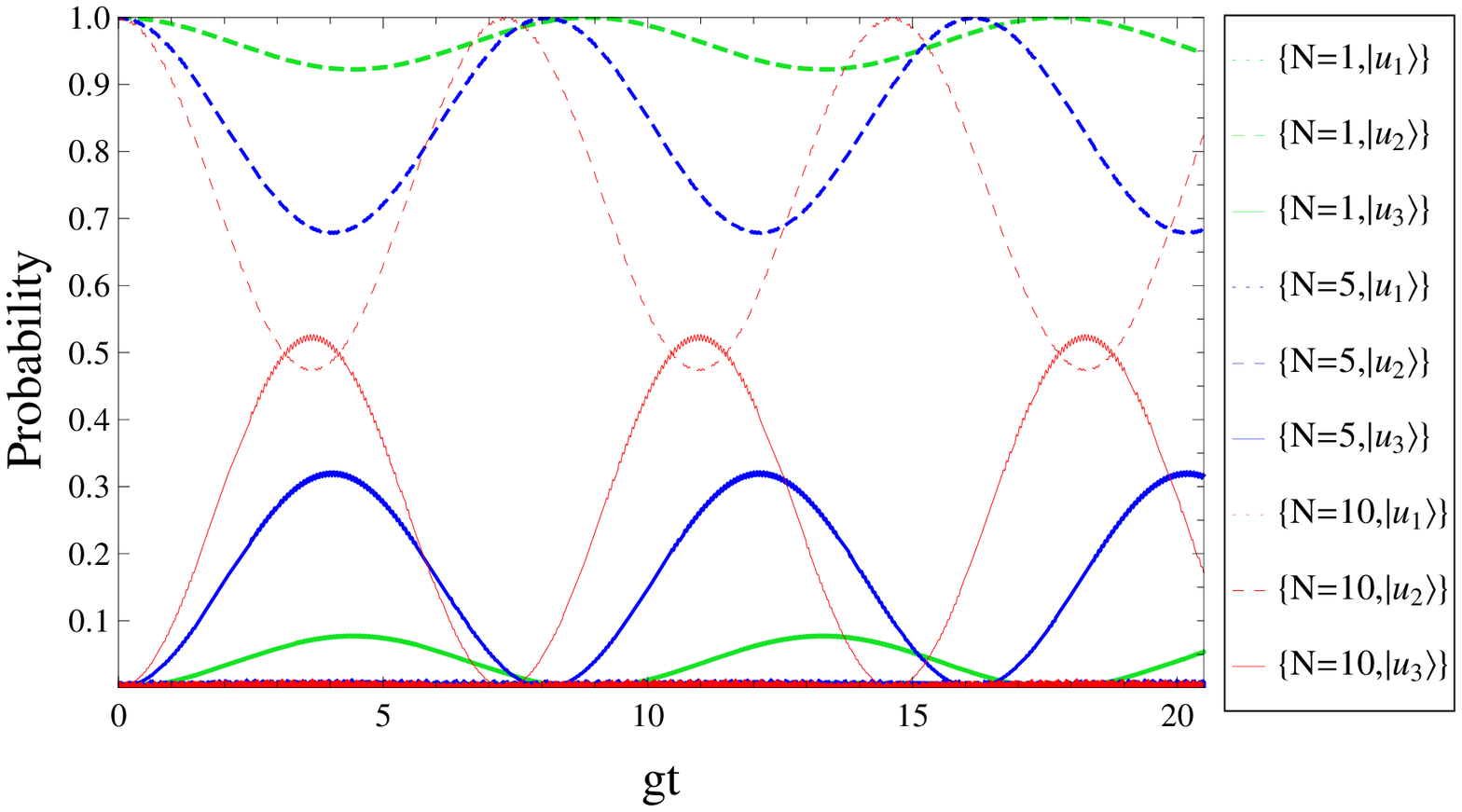}
\caption{\label{fig5}(Color online) Temporal evolution of the  state $\{ \left| {{u}_{1}} \right\rangle =\left| E \right\rangle \left| 0 \right\rangle$, $\left| {{u}_{2}} \right\rangle =\left| G \right\rangle \left| 1 \right\rangle$, $\left| {{u}_{3}} \right\rangle =\left| S \right\rangle \left| 0 \right\rangle \}$ with the initial state in $\left| {{u}_{2}} \right\rangle=\left| G \right\rangle \left| 1 \right\rangle$ considering no dissipations. The solid curves correspond to the state $\left| {{u}_{3}} \right\rangle =\left| S \right\rangle \left| 0 \right\rangle$, the dashed curves correspond to the state $\left| {{u}_{2}} \right\rangle=\left| G \right\rangle \left| 1 \right\rangle$ and the dotted curves correspond to the state $\left| {{u}_{1}} \right\rangle =\left| E \right\rangle \left| 0 \right\rangle$.  With the number of atoms in the cavity growing, there exists more real energy exchanges between the cavity mode and $\left| g \right\rangle \leftrightarrow \left| s \right\rangle $ transition in a certain detuning $\delta$. Due to the large detuning $\Delta$ the probability of the state $\left| {{u}_{1}} \right\rangle =\left| E \right\rangle \left| 0 \right\rangle$ is very small all the time. Other parameters: $\Omega=10g$, $\Delta=100g$ and $\delta=0.3g$.
}
\end{figure}%
In this case if we expect there exist no transitions between the lower states and the excited state, and no real energy exchange between the cavity mode and the $\left| g \right\rangle \leftrightarrow \left| s \right\rangle $ transition, then both $\alpha(t)$ and $\gamma(t)$ should approximate to zero all the time, thus it should be satisfied that $\Delta\gg\sqrt{N}g$, $\delta\gg\sqrt{N}g\Omega/\Delta$. In fact, $\sqrt{N}g\Omega/\Delta$ is the collective Raman Rabi frequency, which should be much smaller than the Raman detuning $\delta$ for preventing the real energy exchanges between the cavity mode and the Raman transition. Similarly, assuming the initial state is $\left| {{u}_{3}} \right\rangle =\left| S \right\rangle \left| 0 \right\rangle $ we can get the condition $\Delta\gg\Omega$ for preventing the transitions between the states $\left| s \right\rangle$ and $\left| e \right\rangle$. Fig.~\ref{fig5} shows that, as the number of atoms involved grows, the Raman transitions between the states $\left| g \right\rangle$ and $\left| s \right\rangle$ happen with higher probability during the state evolution in a certain detuning $\delta$, meaning that there will be notable real energy exchanges between the cavity mode and atoms in the case of large number of atoms involved.

\section{Discussion and Conclusion}
The enhanced coupling strength between the photons and atoms is due to the many-atom interference effects, and actually the many-photon interference effects also exist as the atoms coupled to the field mode. In fact, the coupling strength between the photons and atoms could be enhanced not only by the square root of the number of atoms, but also by the square root of the number of photons involved, thus the large detuning condition is also relevant to the photon number if there are more than one photon participating in the atom-photon interacting. In addition, though all the analyses above are based on the cavity QED system with atoms, the conclusion is also applicable to other kinds of microparticles and artificial microparticles interacting with fields, such as ions, polar molecules and superconducting devices, for the cavity model and free space model.

In conclusion, we have studied the dynamics of the multipartite systems off-resonantly interacting with electromagnetic fields, focusing on the large detuning condition for the effective Hamiltonian in this case. Two-level configuration and three-level lambda configuration under the conditions of two-photon resonance and nonresonace are analyzed theoretically and numerically in detail, and the exact solutions of the temporal evolution of the coupled systems are derived. Since the coupling strength is enhanced due to the many-atom interference effects, we suggest a more rigorous large detuning condition for the effective Hamiltonian in this case. It is significant to apply the rigorous large detuning condition in analyzing the multipartite systems interacting with fields, since the original large detuning condition may lead to the invalidity of the effective Hamiltonian and the errors of the parameters associated with the detuning, especially for the analyses without full numerical simulations.

\section*{Acknowledgments}
This work was supported by the National Natural Science Foundation of China under Grant No. 10774192.

%---------------------------------------------

%---------------------------------------------------------
\end{document}